\pdfoutput=1

\documentclass[twocolumn]{aastex63}
\usepackage{amssymb}
\usepackage{amsthm}
\usepackage{amsmath}
\usepackage{mathrsfs}
\usepackage{scalerel}
\usepackage{graphicx}

\graphicspath{{./}{figures/}}

\begin{document}

\title{Constraining Black Hole Populations in Globular Clusters using Microlensing:\\
Application to Omega Centauri}

\correspondingauthor{John Zaris}
\email{jcz2114@columbia.edu}

\author{John Zaris}
\affiliation{Department of Physics, Columbia University in the City of New York, 550 W 120th St., New York, NY 10027, USA} 

\author{Do\u{g}a Veske}
\affiliation{Department of Physics, Columbia University in the City of New York, 550 W 120th St., New York, NY 10027, USA}

\author{Johan Samsing}
\affiliation{Niels Bohr International Academy, The Niels Bohr Institute, Blegdamsvej 17, DK-2100, Copenhagen Ø, Denmark}

\author{Zsuzsa M\'arka}
\affiliation{Columbia Astrophysics Laboratory, Columbia University in the City of New York, 550 W 120th St., New York, NY 10027, USA}
\author{Imre Bartos}
\affiliation{Department of Physics, University of Florida, PO Box 118440, Gainesville, FL 32611-8440, USA}
\author{Szabolcs M\'arka}
\affiliation{Department of Physics, Columbia University in the City of New York, 550 W 120th St., New York, NY 10027, USA}

\begin{abstract}

We estimate the rate of gravitational microlensing events of cluster stars due to black holes (BHs) in the globular cluster NGC 5139 ($\omega Cen$).  Theory and observations both indicate that $\omega Cen$ may contain thousands of BHs, but their mass spectrum and exact distribution are not well constrained. In this Letter we show that one may observe microlensing events on a timescale of years in $\omega Cen$, and such an event sample can be used to infer the BH distribution. Direct detection of BHs will, in the near future, play a major role in distinguishing binary BH merger channels. Here we explore how gravitational microlensing can be used to put constraints on BH populations in globular clusters.

\end{abstract}


\section{Introduction} \label{sec:intro}

The detection of gravitational waves (GWs) by Advanced LIGO \citep{Aasi_2015} and Advanced Virgo \citep{Acernese_2014} has confirmed the existence of merging binary black holes (BBHs)
\citep{2016PhRvL.116f1102A, 2016PhRvL.116x1103A, 2016PhRvX...6d1015A,
2017PhRvL.118v1101A, 2017PhRvL.119n1101A, 2019arXiv190210331Z, 2019arXiv190407214V}.
However, there is limited evidence to explain how and where this observed BBH population forms in our universe.
The growing list of proposed formation channels includes field binaries
\citep{2012ApJ...759...52D, 2016ApJ...819..108B, 2017ApJ...836...39S, 2017ApJ...845..173M, 2018ApJ...863....7R, 2018ApJ...862L...3S,10.1093/mnras/stz359,10.1093/mnras/sty1999,2017MNRAS.472.2422M},
active galactic nuclei discs
\citep{2017ApJ...835..165B,  2017MNRAS.464..946S, 2017arXiv170207818M, PhysRevLett.123.181101},
galactic nuclei \citep{2009MNRAS.395.2127O, 2015MNRAS.448..754H, 2016ApJ...828...77V, 2016ApJ...831..187A, 2016MNRAS.460.3494S, 2018ApJ...865....2H,2019MNRAS.488...47F},
and dynamical assembly in globular clusters (GCs)
\citep{2000ApJ...528L..17P, 2010MNRAS.402..371B, 2016PhRvD..93h4029R, 2017MNRAS.464L..36A, 2017MNRAS.469.4665P, 2019arXiv190611855A,Ziosi_2014,Mapelli_2016,Di_Carlo_2019,2018PhRvL.121p1103F}. In this work we study
methods to constrain the BH population in GCs independently of GW observations.

\added{Recently, BH candidates have been detected in GCs using a variety of methods, including analysis of X-ray and radio emissions \citep{2012Natur.490...71S,2013ApJ...777...69C,2015MNRAS.453.3918M} and radial velocity measurements of BH companion stars in binary systems \citep{2018MNRAS.475L..15G,2019A&A...632A...3G}.  Stellar-mass BH candidates have even been found in GCs outside of the Milky Way by analyzing X-ray emission patterns \citep{2007Natur.445..183M,2011MNRAS.410.1655M,2010ApJ...721..323S,2010ApJ...725.1805B}.}

Theory and observations indicate that individual GCs are able to retain a large fraction of their initial BH population, depending on their mass and dynamical history \citep[e.g.][]{2015ApJ...800....9M, 2016PhRvD..93h4029R, 2018MNRAS.478.1844A, 2018ApJ...855L..15K, 2019MNRAS.482.4713Z, 2019arXiv191109125W}. One way of probing this population is through GW observations;
but distinguishing BBHs mergers assembled in GCs from those formed through other channels
has been shown to be difficult. Using inferred distributions of BH spins \citep[e.g.][]{2016ApJ...832L...2R},
masses \citep[e.g.][]{2017ApJ...846...82Z}, and orbital eccentricities \citep[e.g.][]{2006ApJ...640..156G, 2014ApJ...784...71S, 2017ApJ...840L..14S, 2018MNRAS.476.1548S, 2018ApJ...853..140S, 2019MNRAS.482...30S, 2018PhRvD..97j3014S, 2018ApJ...855..124S, 2019ApJ...871...91Z, 2019arXiv190711231S, 2019arXiv190905466R} from GW observations is possible; but gives only an
indirect and statistical measure of the contribution of GC BBHs to the set of observed BBH mergers.

In this Letter, we explore the possibility of directly constraining the BH population of GCs located in the Milky Way (MW) through their gravitational lensing effects \citep[e.g.][]{1994AcA....44..165U, 1994AcA....44..235P, 2002ApJ...579..639B}. If BHs populate the core
of GCs, then they will occasionally gravitationally lens and magnify the background cluster stars,
an effect known as microlensing \citep[e.g.][]{1986ApJ...304....1P}.

Previous microlensing studies have investigated several types of lens-source systems.  For example, research has been conducted on the lensing of galactic center stars by GC stars \citep{1994AcA....44..235P, 2012ApJ...744L..18P}, planetary mass objects \citep{2001Natur.411.1022S} and dark matter \citep{1998A&A...336..411J}, and the lensing of GC stars by galactic compact dark matter \citep{1998ApJ...495L..55R} and intermediate-mass BHs theorized to inhabit GCs \citep{2010NewA...15..450S}.

In this work, we study the microlensing of GC stars by stellar-mass GC BHs, focusing our attention on the massive GC $\omega Cen$. Recent studies indicate that a BH population with total mass $\sim 10^5 M_{\odot}$ is likely to occupy the core of $\omega Cen$ \citep{2019MNRAS.482.4713Z}, which makes
this cluster a particularly interesting candidate to monitor in current and future surveys.
Using both analytical and numerical techniques we illustrate that an observable microlensing rate
$\sim 1\ yr^{-1}$ is expected for $\omega Cen$ and investigate how this rate depends on the properties of the BH population.
Any detection or non-detection can therefore be used constrain the current BH distribution in $\omega Cen$.
This, in turn, can help determine the degree to which GCs contribute to the observed BBH merger rate.

The Letter is structured as follows. We begin in Section \ref{sec:theory} by applying microlensing theory to the case of a GC, from which we derive an order-of-magnitude estimate for the lensing rate in $\omega Cen$. In Section \ref{sec:application} we improve on our rate estimate using a more sophisticated Monte-Carlo (MC) technique, where we take into account the observed stellar profile of $\omega Cen$. We conclude our study in Section \ref{sec:con}.

\section{Lensing Theory and Toy Model}\label{sec:theory}

Here, we first review the standard lensing equations \citep[e.g.][]{1986ApJ...304....1P}, which we then use to derive an approximate but closed form expression for the rate of stellar microlensing by BHs in GCs. This expression provides general insight into how the microlensing rate depends on properties such as the mass and velocity dispersion of both the BH and star distributions.

\begin{figure}
    \begin{center}
    \includegraphics[width=\columnwidth]{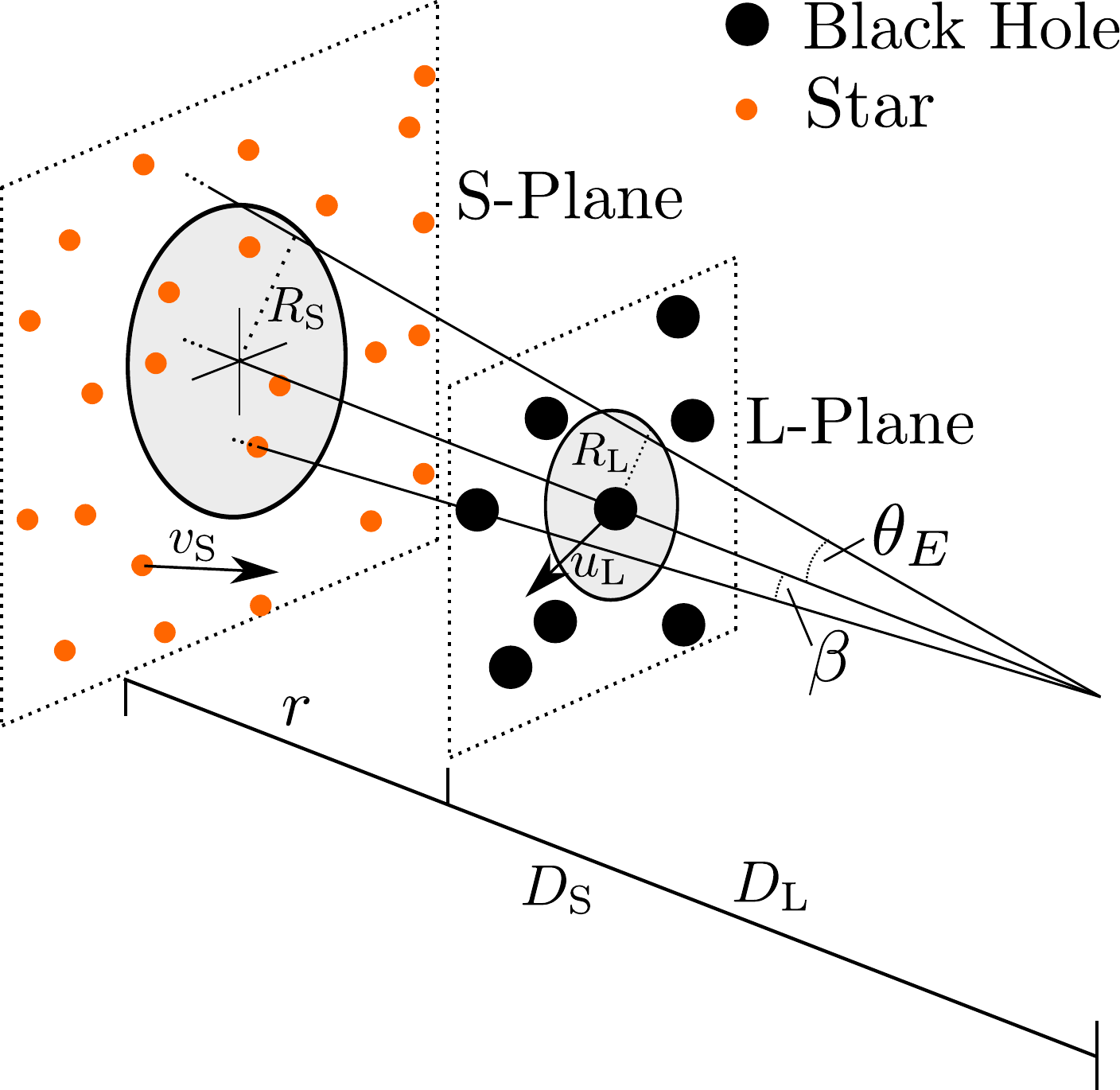}
    \end{center}
    \caption{Illustration of the lensing setup described in Section \ref{sec:theory}. The illustration shows two planes; the source plane ({\it S-Plane}), which here is populated with stars ({\it orange dots}), and the lensing plane ({\it L-Plane}), which is populated with BHs ({\it black dots}). The observer is located to the right at a distance $D_{\rm L}$ and $D_{\rm S}$ to the L-Plane and S-Plane, respectively. The rate at which stars in the S-Plane cross the Einstein ring ({\it grey circle} in the S-Plane with radius $R_{\rm S}$) is linked to the observable microlensing rate, as further described in Section \ref{sec:theory}.}
    \label{fig:toymodel}
\end{figure}

When a lensing object (the BH) passes near the line of sight (l.o.s) from an observer to a source (the star), the
source will appear magnified in the observer's frame by a factor
\begin{equation} 
\mu = \frac{{\alpha}^2+2}{{\alpha}({\alpha}^2+4)^{1/2}}
\label{mag}
\end{equation}
where ${\alpha}$ is the rescaled angular impact parameter defined by
\begin{equation} 
{\alpha} = {\beta}/{\theta_E} 
\end{equation}
In this equation, $\beta$ is the angular distance between the lensed star and the BH, and $\theta_E$ is the angular Einstein radius defined by
\begin{equation} 
\theta_E = \sqrt{\frac{4Gm_{\rm BH}}{c^2}\frac{D_{\rm S}-D_{\rm L}}{D_{\rm S}D_{\rm L}}}
\label{eq:Eang}
\end{equation}
where $D_{\rm S}$ and $D_{\rm L}$ are the distances from the observer to the lensed star and to the BH, respectively, $m_{\rm BH}$ is the mass of the lensing BH, $G$ is Newton's gravitational constant, and $c$ is the speed of light. Figure \ref{fig:toymodel} illustrates this setup.

Using these equations, we now derive an expression for the rate of microlensing in a GC consisting of stars and BHs.
We begin by calculating the rate at which stars located in the source-plane (S-plane)
cross the Einstein ring of a given BH located in the lens-plane (L-plane), where the
radii of the Einstein rings in the L-plane and the S-plane are given by $R_{\rm L} \approx D_{\rm L}\theta_{E}$ and $R_{\rm S} \approx D_{\rm S}\theta_{E}$, respectively. 
Defining $r$ as the distance between the two planes, it follows that the rate at which stars
in the S-plane from $r$ to $r+dr$ pass through the Einstein ring is given by
\begin{equation} 
d\Gamma \approx 2 n(r) R_{\rm S} w dr
\end{equation}
where $n(r)$ is the density of stars in the S-plane at distance $r$, and $w$ is the velocity dispersion of the stars relative to the Einstein ring in the S-plane. Note here that we have ignored the curvature of the S-plane, which is a valid approximation
as $\theta_{E} \ll 1$. The relative velocity dispersion can be expressed as
$w^2 = v_{\rm S}^{2} + (D_{\rm S}/D_{\rm L})^2u_{\rm L}^2$, where $v_{\rm S}$ is the velocity dispersion of the stars in the S-plane and $u_{\rm L}$ is the velocity dispersion of the lensing BHs in the L-plane. Since $D_{\rm L} \gg r$ we have that $R_{\rm S} \approx R_{\rm L}$, and $w^2 \approx v_{\rm S}^{2} + u_{\rm L}^2 \approx u_{\rm L}^{2}$, where the last approximation is accurate within a factor of unity depending on the stellar velocity profile, and on the degree to which the
BHs are in energy equipartition with the stars \citep[e.g.][]{2006ApJ...648..411K, 2013MNRAS.435.3272T}. Generally, the BHs are located near the center of the GC
as they are individually much heavier than the stars. Therefore, their velocity dispersion is
$u_{\rm L} \approx v_{0}$, where $v_{0}$ is the central value.
With these approximations, the differential microlensing rate per BH lens can be written as,
\begin{equation} 
d\Gamma  \approx 2 n(r) D_{\rm L} \theta_{E} v_{0} dr
\label{dgamma}
\end{equation}
Expressing the Einstein angle as $\theta_{E} = \sqrt{2\mathscr{R}r/D_{\rm L}^2}$, where $\mathscr{R}$ is the Schwarzschild radius of a BH with mass $m_{\rm BH}$,  Eq. \eqref{dgamma} can also be written as ${d\Gamma = \sqrt{8} n(r) \sqrt{\mathscr{R}} \sqrt{r} v_0 dr}$. This is the rate for stars in an infinitesimally thin slab located at a distance $(r,r+dr)$ from the L-plane, assuming one BH. Therefore, the total rate for a GC with $N_{\rm BH}$ BHs is given by
\begin{equation} 
\Gamma \approx \sqrt{8} N_{\rm BH} n_{\rm 0} \mathscr{R}^{2} v_{\rm 0} \left( \frac{R_{\rm GC}}{\mathscr{R}} \right)^{3/2} \int n' \sqrt{r'} dr'
\label{zerocurvaturerate}
\end{equation}
where $n' = n/n_0$ is the stellar density scaled by the cluster's central
value, $R_{\rm GC}$ is the radius of the cluster core, $r' = r/R_{\rm GC}$,
and we have assumed that the BHs cluster in the center. As seen, in this simple model we find that $\Gamma \propto N_{\rm BH}{m_{\rm BH}^{1/2}} \propto (N_{\rm BH}m_{\rm BH}){m_{\rm BH}^{-1/2}}$.
Therefore, if the number of BHs is kept fixed $\Gamma \propto {m_{\rm BH}^{1/2}}$,
whereas if the total mass of BHs is kept fixed $\Gamma \propto {m_{\rm BH}^{-1/2}}$.

We can now use Eq. \eqref{zerocurvaturerate} to provide an estimate for the rate of microlensing events in $\omega Cen$. For this we take $N_{\rm BH} = 10^{4}$, $m_{\rm BH} = 10M_{\odot}$ \citep{2019MNRAS.482.4713Z}, \added{$n_0 = 5 \times 10^{4}pc^{-3}$} \citep{2008ApJ...676.1008N,2013MNRAS.429.1887D}, $v_0 = 25\ kms^{-1}$ \citep{2009MNRAS.396.2183S,2010ApJ...719L..60N}, $R_{\rm GC} = 3.25\ pc$ \citep{1995AJ....109..218T,2010arXiv1012.3224H},
the observable threshold of $\mu$ to be $\mu_{\rm obs} = 1.01$ \citep{2017ApJ...842....6B}, $\int n' \sqrt{r'} dr' = 1$ (this integral is $\approx 1$ for most relevant astrophysical profiles), and find \added{$\Gamma \approx 0.2 yr^{-1}$}. This rate is promising and serves as our motivation for exploring this problem in greater detail. We continue below with a more
accurate numerical approach.

\section{Lensing Rate for Omega Centauri}
\label{sec:application}

Having motivated our lensing study of $\omega Cen$ in Section \ref{sec:theory} using analytical arguments, we now move on to
a more accurate model using MC techniques.
Below, we first describe our model of the stars and BHs in $\omega Cen$, after which we present our MC approach and corresponding results.

\subsection{Cluster Model}
\label{sec:bbhej}

Studies of stellar kinematics hint that $\omega Cen$ is likely to harbor a population of BHs with a total mass of $\sim 10^{5} M_{\odot}$ \citep{2019MNRAS.482.4713Z};
however, the BH mass spectrum and distribution are not well constrained. Therefore, to keep our study as model-independent as possible, we adopt the simple `energy equipartition' model from \cite{2006ApJ...648..411K} to describe the radial position and velocity distributions of the BHs, although we note that GCs likely never acquire perfect equipartition \citep[e.g.][]{2013MNRAS.435.3272T}.
In addition, we focus on modeling the microlensing rate from a BH population with a single mass $m_{\rm BH}$ to isolate the mass dependence on our results.
Following \cite{2006ApJ...648..411K}, the BHs uniformly distribute within a sphere of radius
\begin{equation}
\label{core_radius}
R_{\rm BH} = R_{\rm GC} \sqrt{{\langle m\rangle}/{m_{\rm BH}}}
\end{equation}
with a corresponding velocity dispersion of
\begin{equation}
\label{core_velocity}
\sigma_{\rm BH} = \sigma_{\rm GC} \sqrt{{\langle m\rangle}/{m_{\rm BH}}}
\end{equation}
where $\sigma_{\rm GC} \approx \sqrt{({3}/{5}){GM_{\rm GC}}/{R_{\rm GC}}}$,
$M_{\rm GC}$ and $R_{\rm GC}$ are the mass and radius of the cluster core, respectively,
and $\langle m\rangle$ is the mean mass of the GC objects (stars + BHs).

In contrast to the BH population, the stellar distribution in $\omega Cen$ is well constrained from
observations. In this study we use the inferred stellar density and velocity dispersion profiles
from \cite{2013MNRAS.429.1887D} and \cite{2009MNRAS.396.2183S}, respectively.  The former work suggests that the total core mass of $\omega Cen$ is $M_{\rm GC} = 5 \times 10^{5} M_{\odot}$.

\subsection{Monte Carlo method}
\label{sec:mc}

With the two distribution models for the BHs and stars presented above, we are now in a position to
derive the expected microlensing rate for $\omega Cen$. For this, we developed a MC code that operates in the following way.

We first generate a BH assuming that it follows a circular orbit around the center of the core.  The inclination angle of the orbit with respect to the l.o.s is randomized uniformly while the orbital radius and velocity are chosen according to Eqs. \eqref{core_radius} and \eqref{core_velocity}, respectively.  Next, we generate a star whose position and velocity are chosen from the observationally inferred radial density and velocity dispersion profiles,
as described in Section \ref{sec:bbhej}.
At each timestep in the BH's orbit, we then estimate the microlensing magnification $\mu$ of the star and store its maximum value, $\mu_{\rm max}$. This entire process is repeated until a representative sample of star and BH pairs has been simulated.
The final rate can then be calculated by counting the total number of microlensing events per unit
time with $\mu_{\rm max} > \mu_{\rm obs}$, where $\mu_{\rm obs}$ is the observational threshold. \added{For the total rate calculation, we assume a total of $3\times 10^6$ visible stars in the GC. Our simulation also calculates the duration of the lensing events.  Beginning at the maximum brightness magnification, it records the magnification at each subsequent timestep.  From this brightness versus time data, we calculate the minimum time required for the magnification to decrease by the value $\mu_{obs}-1$, which we define as the event duration.  We find that events typically last on the order of several weeks.}

\subsection{Results}
\label{sec:res}

\begin{figure}
    \includegraphics[width=\columnwidth]{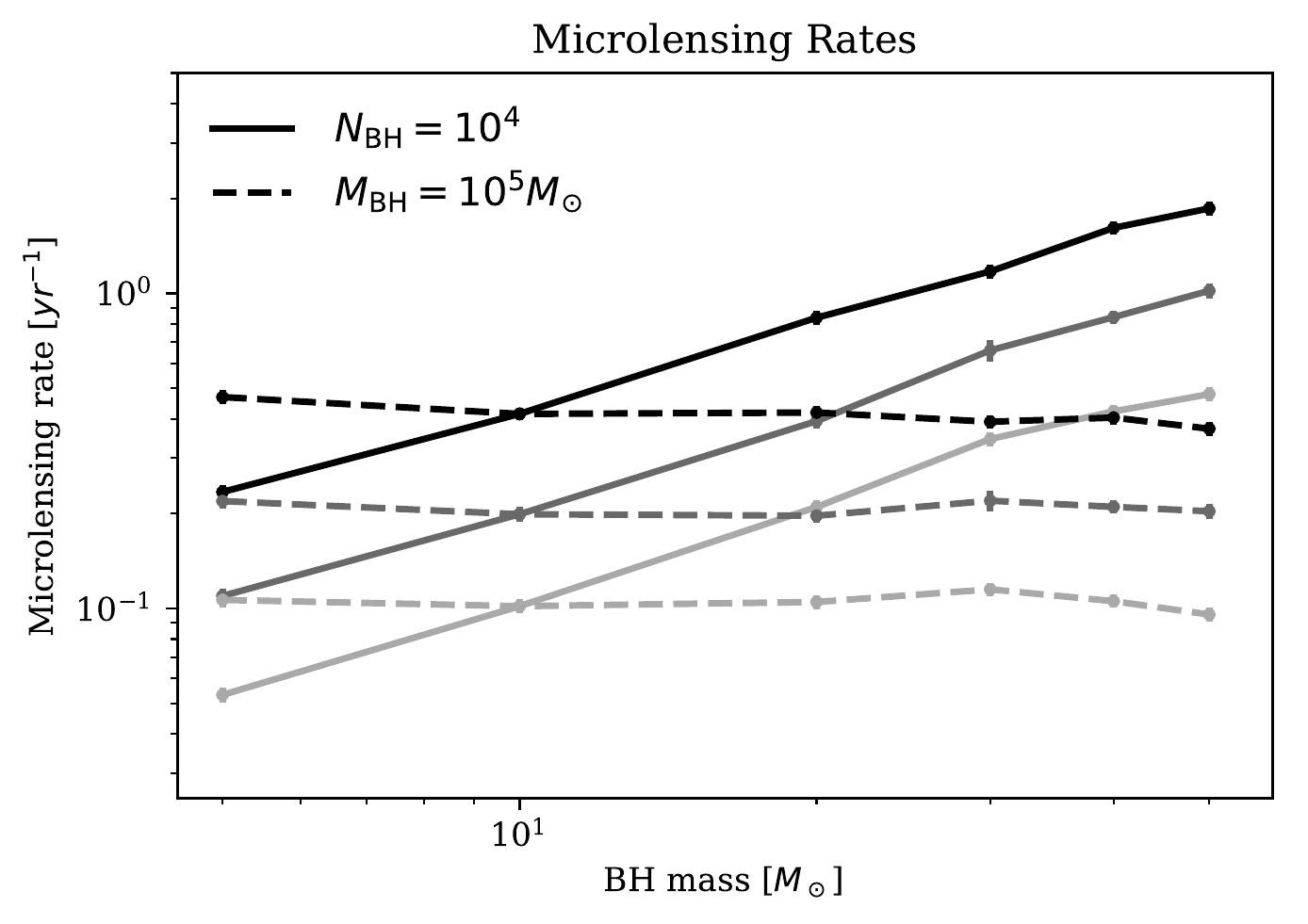}
    \includegraphics[width=\columnwidth]{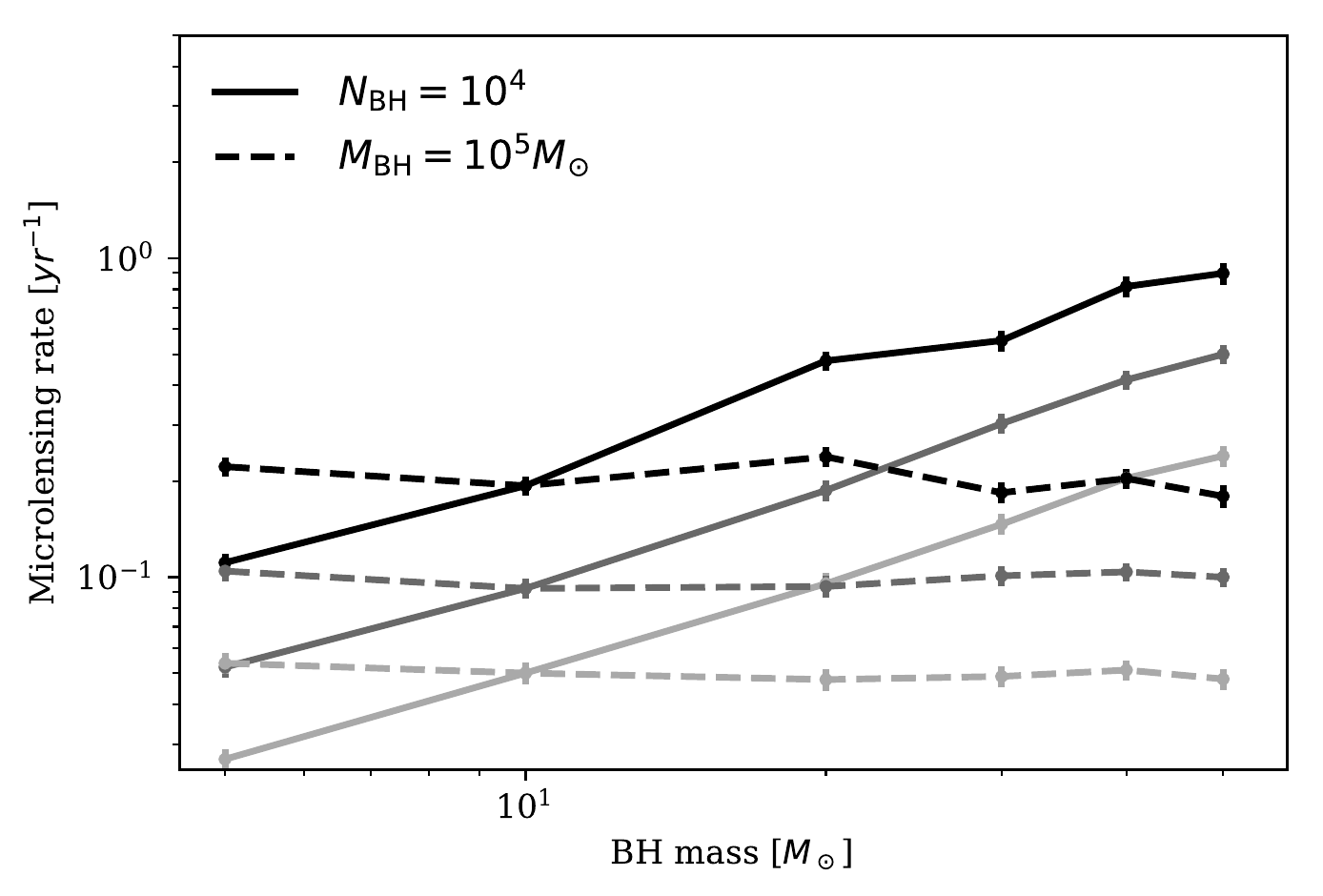}
    \caption{Event rates of cluster stars being microlensed by a corresponding population of cluster BHs in $\omega Cen$. The rates shown are estimated using MC techniques as described in Section \ref{sec:mc}, where the stars and BHs are distributed according to the models outlined in Section \ref{sec:bbhej}. \added{The rates in the top and bottom figures are obtained using magnification thresholds of $\mu_{obs} = 1.01$ and $\mu_{obs} = 1.1$, respectively.} The {\it black, dark gray}, and {\it light gray} lines show results for BH distributions with a radial size of $\{1/2,1,2\} \times R_{\rm BH}$ from Eq. \eqref{core_radius}, respectively. The {\it solid} and {\it dashed} lines show results for when the total number, $N_{\rm BH}$, and total mass, $M_{\rm BH}$, of the BHs are held fixed at $N_{\rm BH} = 10^4$ and $M_{\rm BH} = 10^{5}M_{\odot}$, respectively. Results are discussed in Section \ref{sec:res}.}
    \label{fig:ratevsmass}
\end{figure}

Microlensing rates for $\omega Cen$ derived using our MC simulations described in the above Sections \ref{sec:bbhej} and \ref{sec:mc} are shown in Fig. \ref{fig:ratevsmass}. The solid and dashed lines show results for when the total
number, $N_{\rm BH}$, and total mass, $M_{\rm BH}$, of the BHs are held fixed, respectively. To illustrate the dependence of our results on the uncertain scale of the radial distribution of the BHs, $R_{\rm BH}$, we further show, in differently shaded lines, results for when $R_{\rm BH}$ is varied by a factor of 2 from its fiducial value given by Eq. \eqref{core_radius}. \added{An important parameter is the magnification threshold $\mu_{obs}$, defined as the minimum value of $\mu$ (see \eqref{mag}) associated with an observable brightness magnification.  As seen in Figure 8 of \cite{2017ApJ...842....6B}, the photometric error is smaller for brighter stars, so the magnification threshold is also smaller for brighter stars.  Since we cannot calculate this value for each individual star, we use two different threshold values for the cluster.  From Figure 8 of \cite{2017ApJ...842....6B}, a standard error of 0.1 mags is conservative, as almost all stars have standard errors smaller than this.  This leads to our conservative threshold of $\mu_{obs}=1.1$.  We also calculate the rate for $\mu_{obs}=1.01$, equivalent to a standard error of 0.01 mags.  Note that this is approximately the median standard error from Figure 8.} 

As seen, our numerical results indicate that the expected microlensing rate is in the range $0.1-1\ yr^{-1}$ for $\omega Cen$, which is in good agreement with our analytical results from Section \ref{sec:theory}. For the constant BH number scenario the rate increases slightly faster than \(m_{\rm BH}^{1/2}\), and for the constant total BH mass scenario it decreases slower than \(m_{\rm BH}^{-1/2}\). These behaviors can be explained by the localization of the massive BHs closer to the cluster's center where the star density and, consequently, lensing rate are higher.

\section{Discussion}
\label{sec:con}

Is there a significant population of BHs currently residing in GCs throughout our local volume? That is one of the current major questions in the rising field of GW astrophysics, where merging BHs, but not their origins, are directly observed. As suggested by both theory \citep[e.g.][]{2015ApJ...800....9M, 2016PhRvD..93h4029R} and observations \citep[e.g.][]{2018MNRAS.478.1844A, 2018ApJ...855L..15K, 2019MNRAS.482.4713Z, 2019arXiv191109125W}, GCs are likely able to retain a non-negligible number of BHs; but direct evidence for
BHs in GCs in the upper mass range observed by Advanced LIGO and Advanced Virgo ($\sim 30M_{\odot}$) is still lacking.

In this Letter we have explored the possibility for constraining BH populations in GCs through the use of gravitational microlensing. We find it possible to detect BHs in the core of $\omega Cen$ using microlensing observations with an expected rate range of $\sim 0.1-1\ yr^{-1}$. This rate, each individual lensing lightcurve, and the spatial location of the lensed stars all depend on the BH mass spectrum and distribution. Hence, detections or non-detections of microlensing events can be used to constrain these quantities, although this is not trivial \citep[e.g.][]{1994AcA....44..165U, 1994AcA....44..235P}. \added{Furthermore, while we are concerned only with microlensing events in which a BH acts as the lens for a cluster star, it is also possible for another cluster star to serve as the lens.  Observationally, these two cases are distinguishable.  A star lens can be observed optically while a BH lensing event is characterized by an unobservable lens.} 

This strategy described in this Letter is naturally not limited to $\omega Cen$, but can be applied to any of the $\sim 150$ GCs in the MW. \added{However, it is important to keep in mind that $\omega Cen$ is a unique GC which likely has an unusually high microlensing rate.  $\omega Cen$ has a greater mass than other Galactic GCs (e.g. \cite{2018MNRAS.478.1520B}), and has even been proposed to be a tidally stripped dwarf galaxy \citep{2000LIACo..35..619M,2019NatAs...3..667I}.  Additionally, it cannot be assumed that all GCs contain as many BHs as $\omega Cen$.  However, several clusters analyzed in recent works likely contain hundreds of BHs and, therefore, may have significant microlensing rates \citep{2018MNRAS.479.4652A,2018MNRAS.478.1844A, 2018ApJ...855L..15K, 2019arXiv191109125W}}.

Observing BH microlesning events in GCs is challenging, but will likely soon become possible as telescopes with improved performance continue to be constructed. This includes both ground-based telescopes, such as the Thirty Meter Telescope, and
space-based ones, such as The Wide Field Infrared Survey Telescope (WFIRST). Past studies of data from the Hubble Space Telescope have already been used to analyze microlensing events near the galactic center and have successfully constrained lens masses \citep[e.g.][]{2017ApJ...843..145K}. Determining the mass of GC BHs may be more difficult as the source and lens distances are similar and, therefore, must be measured precisely.
\added{Our simulations suggest that microlensing events last on the order of several weeks.  Therefore, observing the cluster on the order of once every few days should provide sufficient data to capture most lensing events.}
Even though resolution continues to improve, observations near the GC core center may still face the issue of crowding, in which it is impossible to resolve two nearby stars. In this case, lensing will still be observable,
but since the localization of the lensed stars can be ambiguous, the uncertainty of the inferred BH population parameters will be higher. In follow up work we will
study how to optimize current and future search strategies for observing such BH microlensing events.

\section{Acknowledgements}
The authors thank No\'e Kains for discussions. The authors appreciate the generous support of Columbia University in the City of New York and the University of Florida. JZ is grateful for support from the Ng Teng Fong Student Internship Fund administered through the Columbia College Summer Fund. JS acknowledges support from the Lyman Spitzer Fellowship,
and funding from the European Union’s Horizon 2020 research
and innovation programme under the Marie Sklodowska-Curie grant agreement No. 844629.
DV is grateful to the Ph.D. grant of the Fulbright foreign student program.

\bibliography{sample63}{}
\bibliographystyle{aasjournal}

\end{document}